\documentclass[journal,authorversion,nonacm]{acmart}

\AtBeginDocument{%
  \providecommand\BibTeX{{%
    \normalfont B\kern-0.5em{\scshape i\kern-0.25em b}\kern-0.8em\TeX}}}

\setcopyright{acmcopyright}
\copyrightyear{2022}
\acmYear{2022}

\acmConference[ACM CHI '22 - VR $\lbrack we$ $are \rbrack$ Training Workshop]{Make sure to enter the correct
  conference title from your rights confirmation email}{April 25, 2022}{online}
\begin{document}

\title[MED1stMR: Mixed Reality to Enhance Training of Medical First Responder]{MED1stMR: Mixed Reality to Enhance the Training of Medical First Responders for Challenging Contexts}

\author{Helmut Schrom-Feiertag}
\email{helmut.schrom-feiertag@ait.ac.at}
\orcid{}
\author{Georg Regal}
\email{georg.regal@ait.ac.at}
\orcid{}
\author{Markus Murtinger}
\email{markus.murtinger@ait.ac.at}
\orcid{}
\affiliation{%
  \institution{AIT Austrian Institute of Technology}
  \streetaddress{Giefinggasse 4}
  \city{Vienna}
  \state{Austria}
  \country{Vienna}
  \postcode{1210}
}


\renewcommand{\shortauthors}{Schrom-Feiertag, Regal and Murtinger}

\maketitle

\section{Introduction}

Mass-casualty incidents with a large number of injured persons caused by human-made or by natural disasters are increasing globally. In such situations, medical first responders (MFRs) need to perform diagnosis, basic life support, or other first aid to help stabilize victims and keep them alive to wait for the arrival of further support. Situational awareness and effective coping with acute stressors are essential \cite{frenkel2021impact} to enable first responders to take appropriate action that saves lives. Such tasks are particularly challenging for first responders, that lack the special training to act optimally in these situations. Increasingly severe consequences of natural disasters and terrorist threats will expand the occurrence probability of such stressful and demanding situations and require the development and deployment of innovative technological solutions adapted to the (cross-sectoral) needs of first responders.

In that context, the adage “practice makes perfect” is well-fitting to situational training. Lectures, books, videos, etc. are no substitute for hands-on experiences, and humans often learn more from their mistakes than from their successes. Unfortunately, it is difficult to provide such training for large-scale emergency medicine or in dangerous conditions. Current training of medical first responders often happens through live exercises: medics practice stabilisation and wound care via moulage, a training exercise where live persons are given highly realistic “fake” wounds. The drawback of such training is the large effort needed to create such live training exercises (a large number of ‘victim actors’ needed, availability of infrastructure, etc.) and the lack of realistic treatments. Hence, such training or exercises are not executed very often and sometimes also fail to create “real” stressful and demanding environments. 

Virtual Reality (VR) has already been demonstrated in several domains to be a serious alternative, and in some areas also a significant improvement to conventional learning and training. 
Especially for the challenges in the training of MFRs, it can be highly useful for practising and learning domains where the context of the training is not easily available. VR training offers controlled, easy-to-create environments that can be created and trained repeatedly under the same conditions. This repetition makes it possible to master a new skill or process. Like in real-life training, trainees are transformed into active users who need to be physically and mentally engaged to evaluate the situation, take appropriate measures and act accordingly. 

There are two types of VR medical training systems realised up to date. Systems centred on teaching direct physical skills and procedures e.g. surgery (e.g. \cite{harrington2018development}, \cite{jaskiewicz_applicability_2019}, \cite{nickel2015virtual}, \cite{kenngott2018mobile}). Those usually employ highly sophisticated hardware user interfaces providing realistic haptic feedback or/and mimicking real devices. On the other end of the spectrum, there are VR systems helping the trainee to develop psychological skills required in real-world scenarios  (e.g. \cite{pallavicini2016virtual}, \cite{kaplan2017role}) and in particular decision training. 

However, these approaches are currently rather disjunct and do not allow to train decision-making under stress together with physical skill scenarios. Also, other medical tasks or preclinical routines in patient care and treatment are not yet covered by haptic solutions. Training medical skills are about vision and haptics for tangible interaction, and if a simulation has only one of those two, it will provide only half of the experience. 

\section{Vision of the project MED1stMR}

As an advanced alternative to VR, Mixed Reality (MR) environments have the potential to augment current VR training by providing a dynamic simulation of an environment and hands-on practice on injured victims. 

There are several interpretations of MR, one is that the real environment is augmented by digital objects, another is that physical objects are integrated into the VR.
In our vision, MR is to be considered as the latter interpretation with a fully digital environment, where the user sees a fully digital environment without looking at the real world, but this digital environment is connected to real physical objects. This VR-based mixed reality is also called augmented virtuality (AV) according to \cite{milgram1995augmented}. 

Building on this interpretation of MR, the main aim of MED1stMR is to develop a new generation of MR training with haptic feedback for enhanced realism. To this end, we will pursue the following pioneering concepts: 

\subsection{Integration of high-fidelity patient simulation manikins for enhanced realism}
Evidence shows that the use of VR is useful when the training domain is complex and difficult to master (cf. \cite{huang2010investigating}, \cite{bossard2008transfer}) and when the audio-visual features assisted by haptic feedback of the training environment are crucial to the overall training success (cf. \cite{abate2009haptic}). This makes virtual environments the solution for practising and learning domains where the context of the training is not easily available or replicable due to security and safety issues (e.g. \cite{grabowski2015virtual}). It allows creating easily a diverse range of training scenarios tailored to the training goals and needs (single user vs. teams, from single to many people injured in large incidents, influence of psychological and contextual factors).

Through the integration of high-fidelity patient simulation manikins and medical equipment into the MR experience, MED1stMR offers a much richer sensory experience. This MR training environment allows trainees to immerse into virtual scenarios and be able to feel and perceive actual movements of the limbs, head, and face through tactile and visual interaction as they are actuated. Furthermore, it enables systematic manipulation of a large set of potential influence factors in order to optimise training effects. This will bring virtual training closer to reality and enable both scenario training and medical training in the same MR training environment. 


\subsection{Biosignal feedback loop and smart scenario control to enhance effectiveness of MR training}
The wireless integration of wearables in MR training environments is emerging (e.g., \cite{havard2020meta}) and particularly in the training of highly demanding skills such as piloting/aviation, medical surgeries (e.g., \cite{currie2019wearable}) or first responders (e.g., \cite{carroll2020automatic}). 

In order to better support, assist and personalise MFRs training, we will integrate wearable technology for monitoring trainees’ physiological data. Smart electronic devices can detect and transmit information regarding biosignals, informing on trainees’ physiological status. Monitoring these signals will provide the detection of physical and psychological strain and stress during training. 

This will provide information for the debriefing sessions and can be used for real-time scenario control through the trainer (manual control) or automatically by the training system through artificial intelligence-based adaptive smart scenarios. The data on trainee state and behaviour can then be used to constitute a feedback loop for personalising and adapting training to the trainees' needs. Such a system can automatically adjust the scenarios according to the stress level of the trainee, for example, low stress increases the difficulty of the scenario and allows longer and more complex scenarios to be trained without the intervention of a trainer.

\section{Our Motivation}

The development of such a training system for MFRs requires research, expertise, and knowledge in the areas of medical research, biosensors, and wearable technologies, human factors research, psychology, physiological research, technology experience, user research, VR/MR, and medical training simulation development to answer all the questions that arise in order to develop an optimal training system for MFRs:

\begin{itemize}
    \item How can haptic feedback for training medical skills on victims be provided for MFRs?
    \item Which scenarios and use cases are most suitable for MR training and deliver the greatest benefits?
    \item How can effective MR training scenarios be developed?
    \item How effective are such training approaches, how good is the learning progress and how does it compare to real-world training?
    \item How should a MR training curriculum be designed and merged with existing training curricula?
    \item What about the costs for the training system and does the benefit justify the effort?
\end{itemize}


\section{Our contribution to the workshop}

MED1stMR develop a MR training system based on a combination of VR environment and the integration of VR-enabled manikins. With this new training environment, MED1stMR delivers a training platform for collaborative multi-user training to train the medical skills of MFRs as well the decision-making abilities in disaster situations. In the project, the training is designed for teams of up to four people to enable emergency teams to train together. The technological basis is the Refense trainings platform (\url{www.refense.com}) that allows up to 10 users on an area of 11 x 20 meters to immerse themselves into a realistic common shared scenario, see each other in real time with full-body VR tracking. A trainer can also be included as an invisible observer in the training. Every movement and voice spoken are recorded for debriefing of team collaboration and actions taken.
The inclusion of manikins as tangible objects as learning support provides a more realistic experience and enables novel possibilities for hands-on tasks. The basis will build the ADAM-X manikin (\url{https://medical-x.com/product/adam-x/}) and will be advanced to a fully functional touch-enabled human manikin designed for practising skills in trauma emergency situation.

The goal of the MED1stMR training solution is to train the situation awareness and the procedure in the first and second triage. The realisation of a biosignal feedback loop with body sensors allows to monitor trainee (stress, anxiety, etc.) states and behaviours of MFRs during training and will make this data available for scenario control. For this purpose, heart rate variability is measured as one of the most reliable indicators of stress. The trainer can adapt training to the personal needs of trainees and provides a new way of interaction between trainer and trainee and requires an appropriate user interface for support.

A key point in MED1stMR is to examine the effectiveness of training for the different roles. To increase the effectiveness of the training, the simple repetition of the training scenarios as well as the recording of all movements, activities and communication during the training for the debriefing play an important role. The presentation of our project is intended to provide an insight into our approach and to open up an exchange of experiences with other people, projects, research, and developments and also to get an impression of how the topics are received, what others are doing in this field and where further and interesting research topics lie in this area.

We can contribute existing knowledge to the workshop and discuss with the other participants' challenges and help to set up a future research agenda for collaborative multi-user VR training.

\section*{Acknowledgements}
The project MED1stMR has received funding from the European Union’s Horizon 2020 Research and Innovation Programme under grant agreement No 101021775. The content reflects only the MED1stMR consortium's view. Research Executive Agency and European Commission is not liable for any use that may be made of the information contained herein.

\bibliographystyle{ACM-Reference-Format}
\bibliography{sample-base}


\end{document}